\documentclass[a4paper,11pt]{article}
\usepackage{pos}

\title{Vector bosons and jets at the LHC}

\author*[a]{Mathieu Pellen}

\affiliation[a]{University of Cambridge, Cavendish Laboratory,\\
  19 JJ Thomson Avenue, Cambridge CB3 0HE, United Kingdom}

\emailAdd{mpellen@hep.phy.cam.ac.uk}

\abstract{
In these proceedings I review recent theoretical predictions describing processes involving vector bosons and jets at the LHC.
Such processes possess very distinctive phenomenologies and their theoretical accuracy is very different.
I focus here on three illustrative cases:
the production of a Z boson in association with a b jet, W-pair production in association with a jet, and vector-boson scattering.
Some of these theoretical results are also compared to experimental data.
}

\FullConference{%
  The Eighth Annual Conference on Large Hadron Collider Physics-LHCP2020 \\
  25-30 May, 2020\\
  online}

\begin{document}
\maketitle

In this short proceedings three computations involving gauge boson and jets at the LHC are presented:
the production of a Z boson in association with a b jet, W-pair production in association with a jet, and vector-boson scattering.
The accuracy and the phenomenology of such processes are very different and therefore offer a good sample of current state-of-the-art theoretical predictions.
I would like to emphasis that the results reviewed here are not original and have already been presented elsewhere.
Also, I am not author of all the results presented here and therefore refer the interested reader to the original references provided.

\section{Z+b~jet}

 \begin{figure}
   \begin{center}
     \includegraphics[width=7cm]{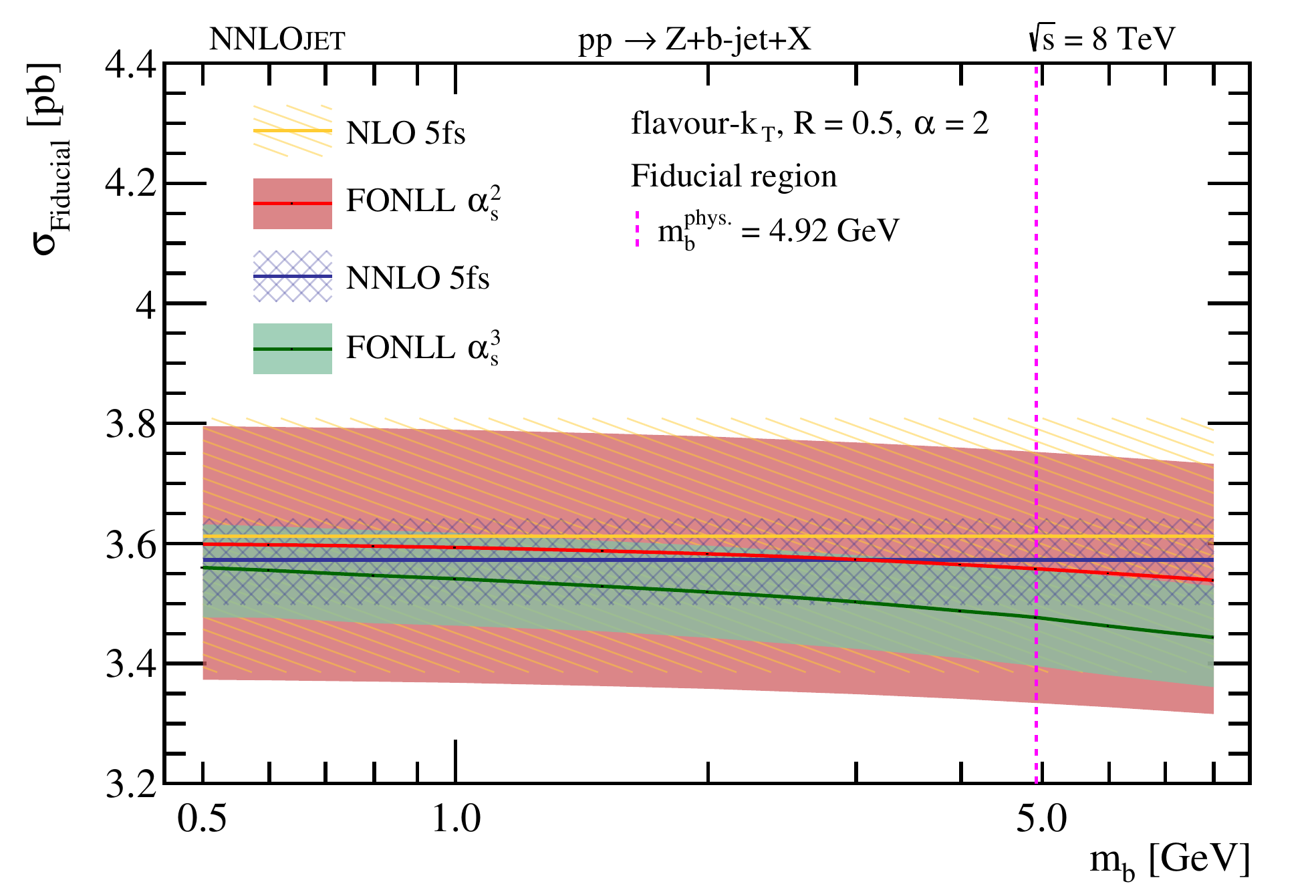}
     \includegraphics[width=7cm]{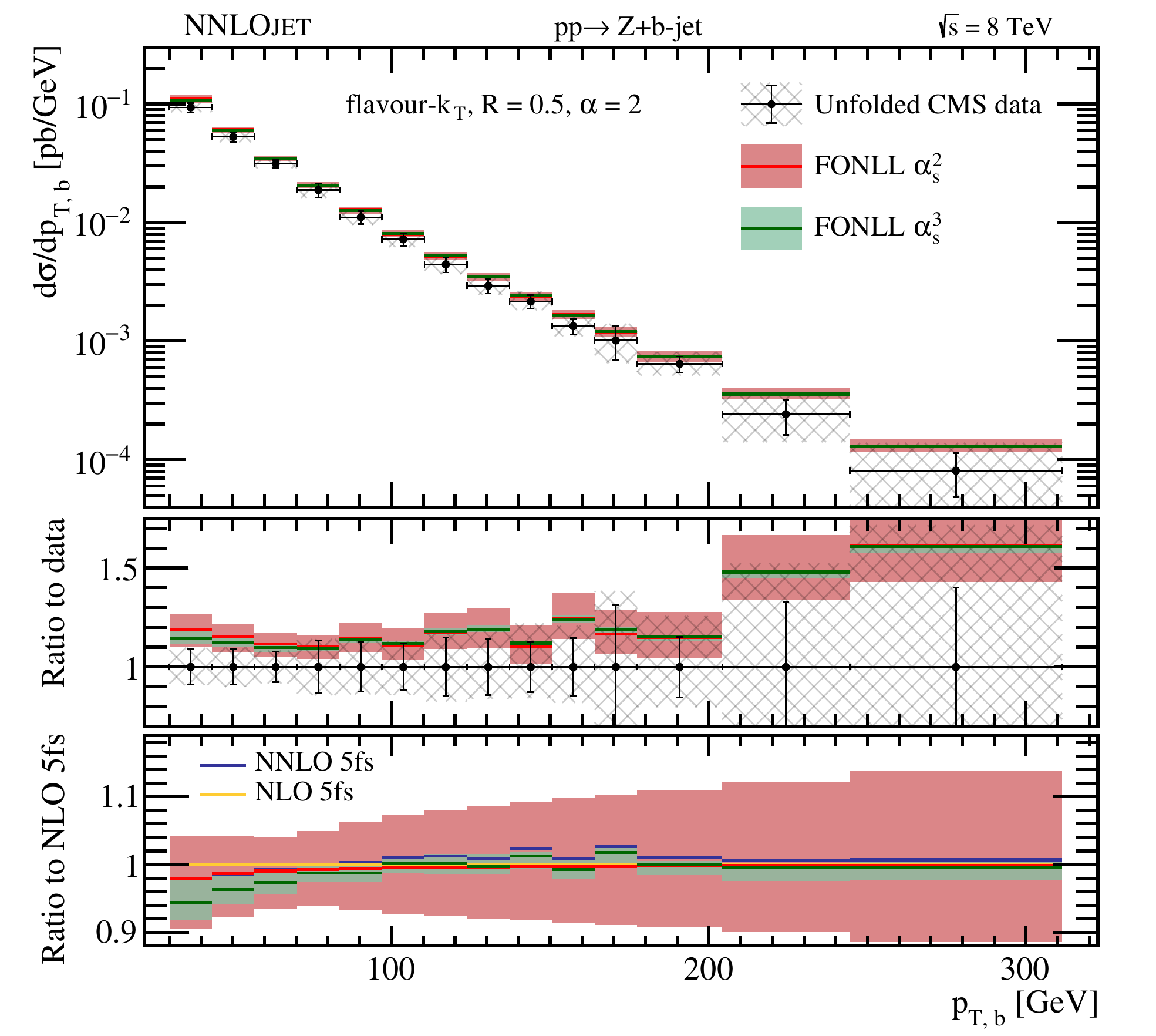}
   \end{center}
   \caption{
These figures are taken from Ref.~\cite{Gauld:2020deh} and are for ${\rm p}{\rm p} \to {\rm Z}+{\rm b}$ at the LHC.
Left: Cross section as a function of the bottom-quark mass for various theoretical predictions.
Right: Various theoretical predictions compared to experimental data for the differential distribution in the transverse momentum of the bottom jet.
}
\label{zb}
 \end{figure}

Reference~\cite{Gauld:2020deh} presents a new NNLO QCD computation for the production of a Z boson in association with a b jet including bottom-quark mass effects.
It is made of a NNLO-QCD computation in the 5 flavour combined with NLO QCD corrections in the 4 flavour at order $\mathcal{O}\left(\alpha_{\rm s}^3 \alpha^2\right)$.
Schematically it proceeds as  ${\rm d}\sigma^{\rm FONLL} = {\rm d}\sigma^{\rm 5fs} + \left( {\rm d}\sigma_{m_b}^{\rm 4fs} - {\rm d}\sigma^{\rm 4fs}_{m_b\to0} \right)$ which allows to incorporate exact b-mass effects.
The dependence on the bottom-quark mass is shown in Fig.~\ref{zb} (left).
One interesting aspect of this computation is that it features a flavoured jet in the final state, therefore requiring the use of the flavoured $k_{\rm T}$ algorithm \cite{Banfi:2006hf} in order to account for soft wide-angle ${\rm q}\bar{\rm q}$ pairs in a infrared-safe way.
On the experimental side, jets are reconstructed with the anti-$k_{\rm T}$ algorithm and only then the flavour of the jets is identified.
Applying non-perturbative corrections to account for this effects to the CMS data \cite{Khachatryan:2016iob} raises: 
$\sigma_{\rm Fiducial,f\text{-}k_{}\rm T}^{\rm CMS} = 3.119\pm0.212^{+0.021}_{-0.032}{\rm pb}$.
The theoretical predictions at order $\mathcal{O}(\alpha_s^3)$ gives $\sigma^{\rm FONLL}_{\rm Fiducial}(m^{\rm phys.}_b) = 3.477^{+0.081}_{-0.081}{\rm (scales)}~{\rm pb}$ which is in good agreement with the measurement.
The same behaviour can be observed at the level of the differential distribution in the transverse momentum of the bottom jet in Fig.~\ref{zb} (right).

\section{WW+jet}

In Ref.~\cite{Aaboud:2016mrt}, the first measurement of ${\rm p}{\rm p} \to {\rm W}^+ {\rm W}^- {\rm j}$ at the LHC has been presented by the ATLAS collaboration.
One interesting aspect of this study is the combined analysis of ${\rm p}{\rm p} \to {\rm W}^+ {\rm W}^- {\rm j}$ and ${\rm p}{\rm p} \to {\rm W}^+ {\rm W}^- {\rm j}$.
In particular ratios of the two processes have been presented there.
In Ref.~\cite{Brauer:2020kfv}, theoretical predictions to address such a measurement have been presented.
They have been obtained with the fully automated framework {\sc Sherpa}+{\sc Recola} \cite{Bothmann:2019yzt,Actis:2016mpe,Biedermann:2017yoi}.

\begin{figure}
   \begin{center}
     \includegraphics[width=7cm]{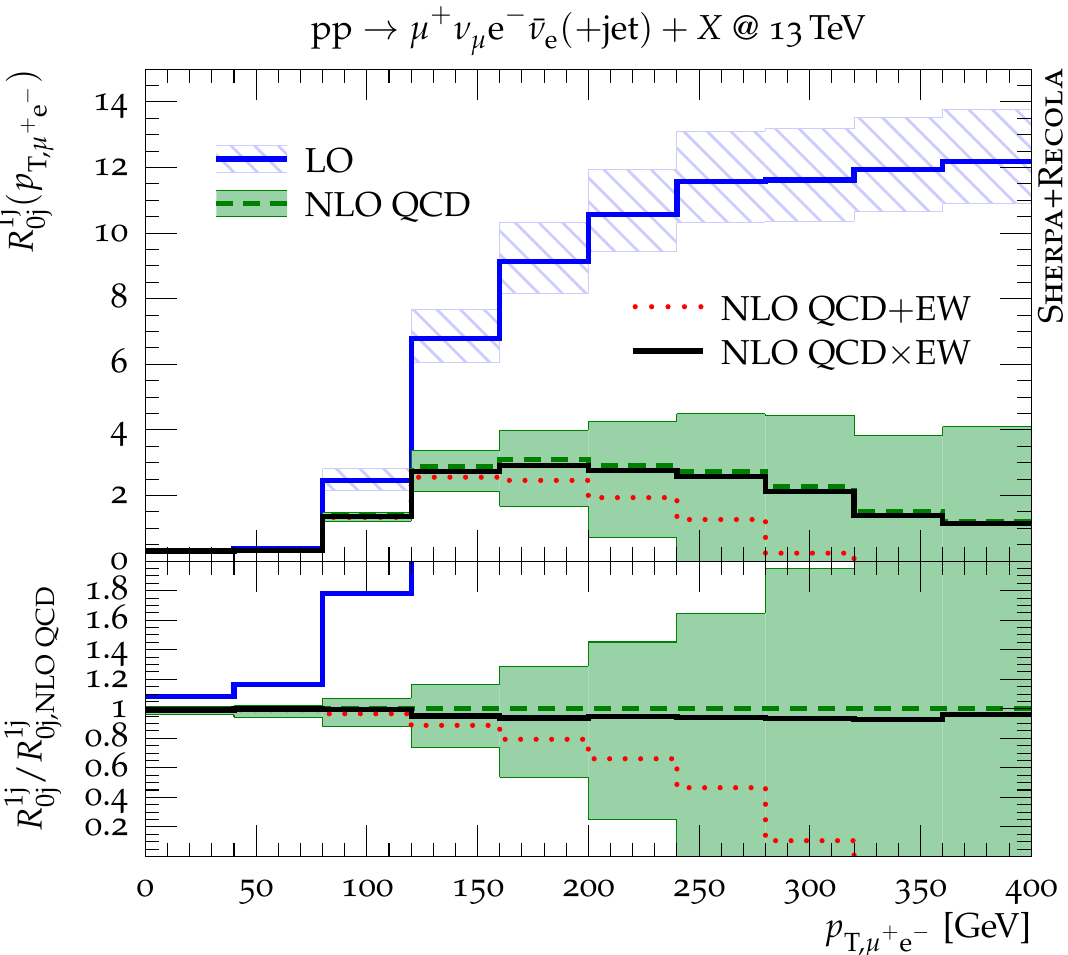}
     \includegraphics[width=7cm]{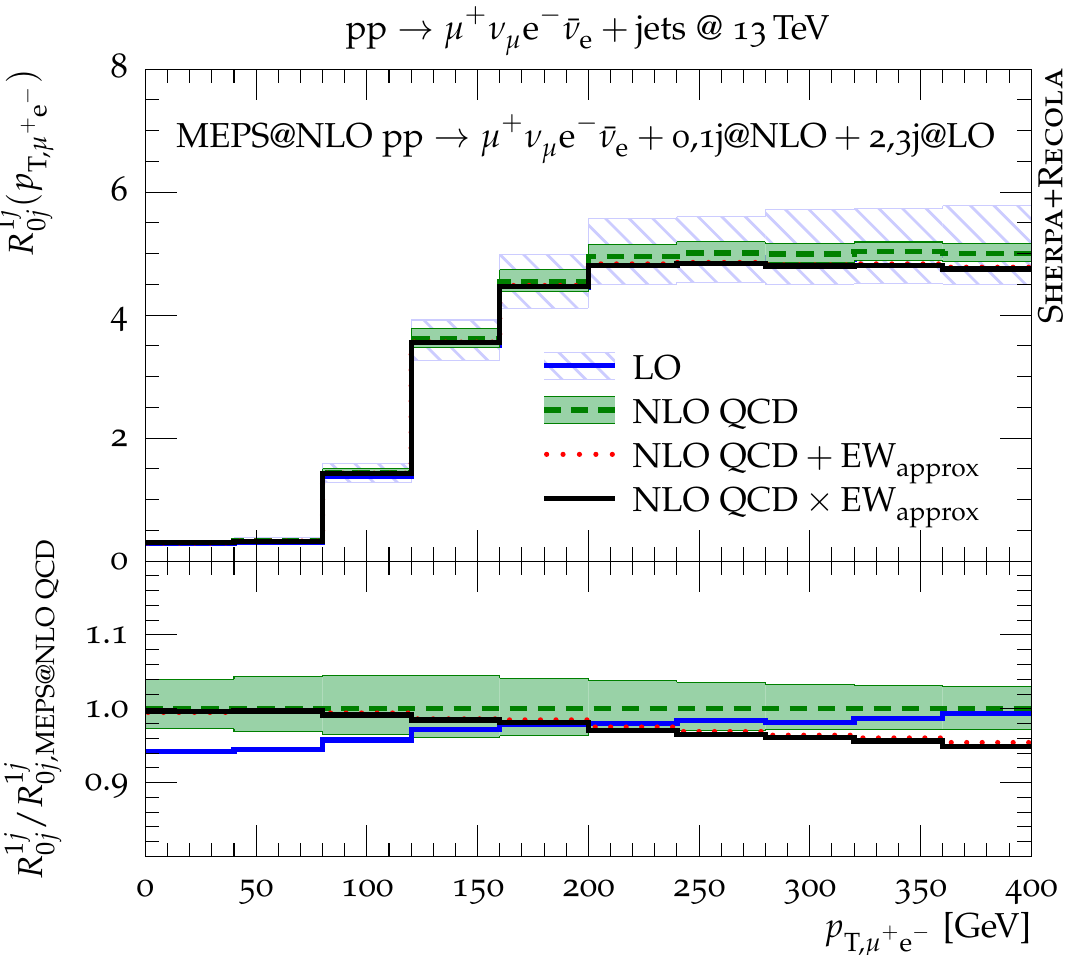}
   \end{center}
   \caption{
These figures are taken from Ref.~\cite{Brauer:2020kfv}.
Ratio plots between ${\rm p}{\rm p} \to {\rm W}^+{\rm W}^- {\rm j}$ and ${\rm p}{\rm p} \to {\rm W}^+ {\rm W}^-$ for
the transverse momentum of the two charged leptons at fixed order (left) and including merging of different multiplicities with parton shower effects (right).
}
   \label{wwj}
 \end{figure}

The first aspect of this publication is the fixed-order analysis at NLO QCD+EW for the off-shell processes ${\rm p}{\rm p} \to \mu^+ \nu_\mu {\rm e}^- \bar \nu_{\rm e}$ and ${\rm p}{\rm p} \to \mu^+ \nu_\mu {\rm e}^- \bar \nu_{\rm e} {\rm j}$ at the LHC.
The second aspect of this work is the production of an inclusive sample featuring merged predictions including parton-shower effects and approximate electroweak corrections: ${\rm p}{\rm p} \to \mu^{ +} \nu_\mu {\rm e}^- \bar \nu_{\rm e} + 0,1 {\rm j} {\rm @NLO} + 2,3 {\rm j} {\rm @LO} $.
With this sample at hand, one can then study theoretical predictions for different jet multiplicities ($n_j = 0$ and $n_j = 1$) which can then be compared to the fixed-order predictions.
Note that in all theoretical predictions presented in Ref.~\cite{Brauer:2020kfv}, a jet veto has been applied following the experimental set-up.
In Fig.~\ref{wwj}, the ratio of the two processes as a functions of the transverse momentum of the two charged leptons is shown.
It is interesting to observe that merged predictions (right) provide much more stable ratios with respect to higher-order corrections than the fixed-order ones (left).

\section{Vector-boson scattering}

Vector-boson scattering at the LHC is characterised by the scattering of heavy gauge bosons radiated off two quark lines.
Its signature is thus two gauge bosons in association with two jets at order $\mathcal{O}\left({\alpha}^6 \right)$.
The exact same signature also exists at order $\mathcal{O}\left({\alpha_{\rm s}} {\alpha}^5 \right)$ (interference contribution) and $\mathcal{O}\left({\alpha_{\rm s}}^2 {\alpha}^4 \right)$ (QCD contribution) which constitute irreducible backgrounds.
At NLO, there is thus a tower of four NLO contributions appearing and some of them are mixed contributions making thus the distinction between the EW and QCD component meaningless \cite{Biedermann:2017bss}.

\begin{figure}
\begin{center}
 \includegraphics[width=7cm]{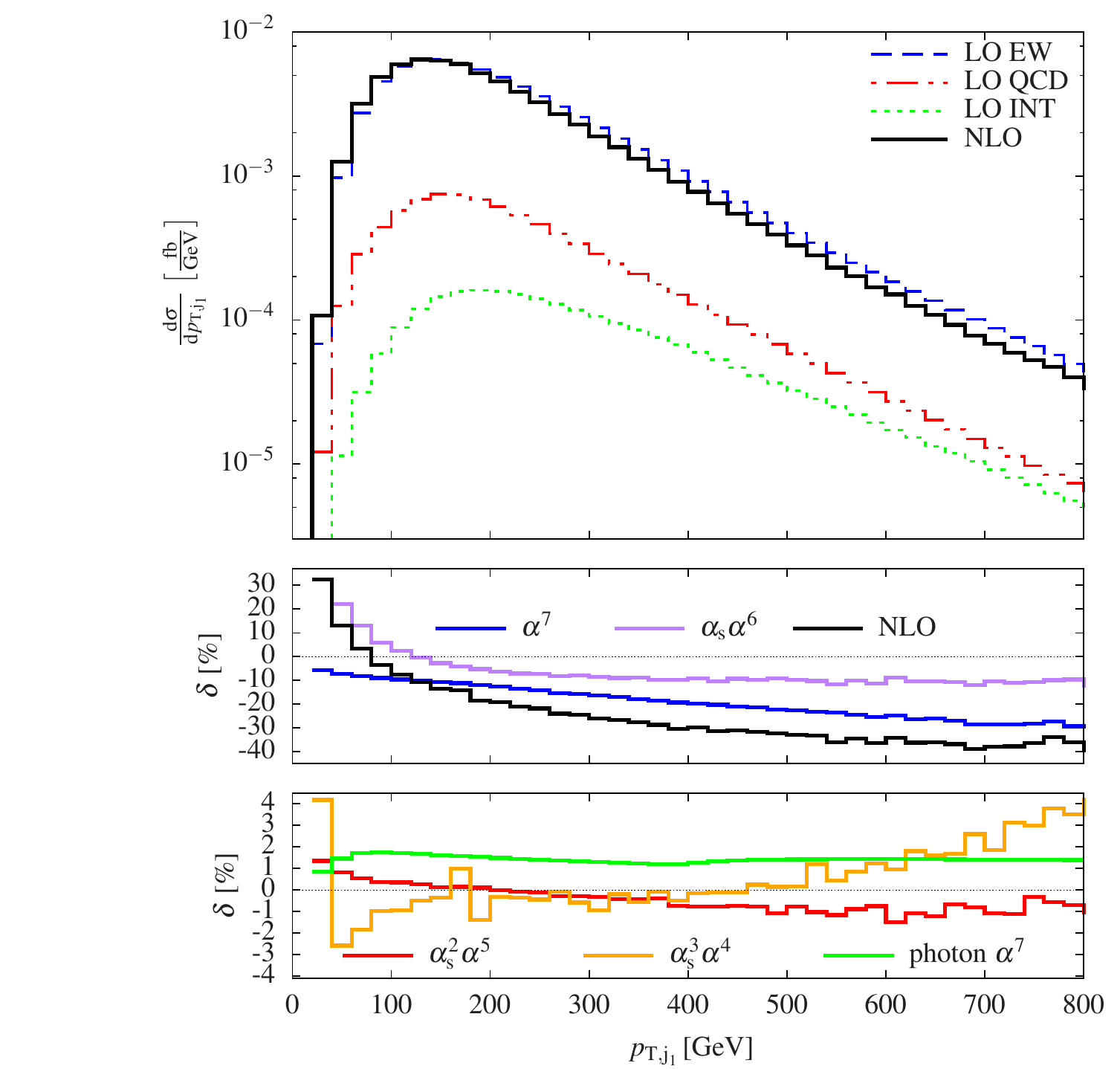}
 \raisebox{-0.4cm}{\includegraphics[width=7.8cm]{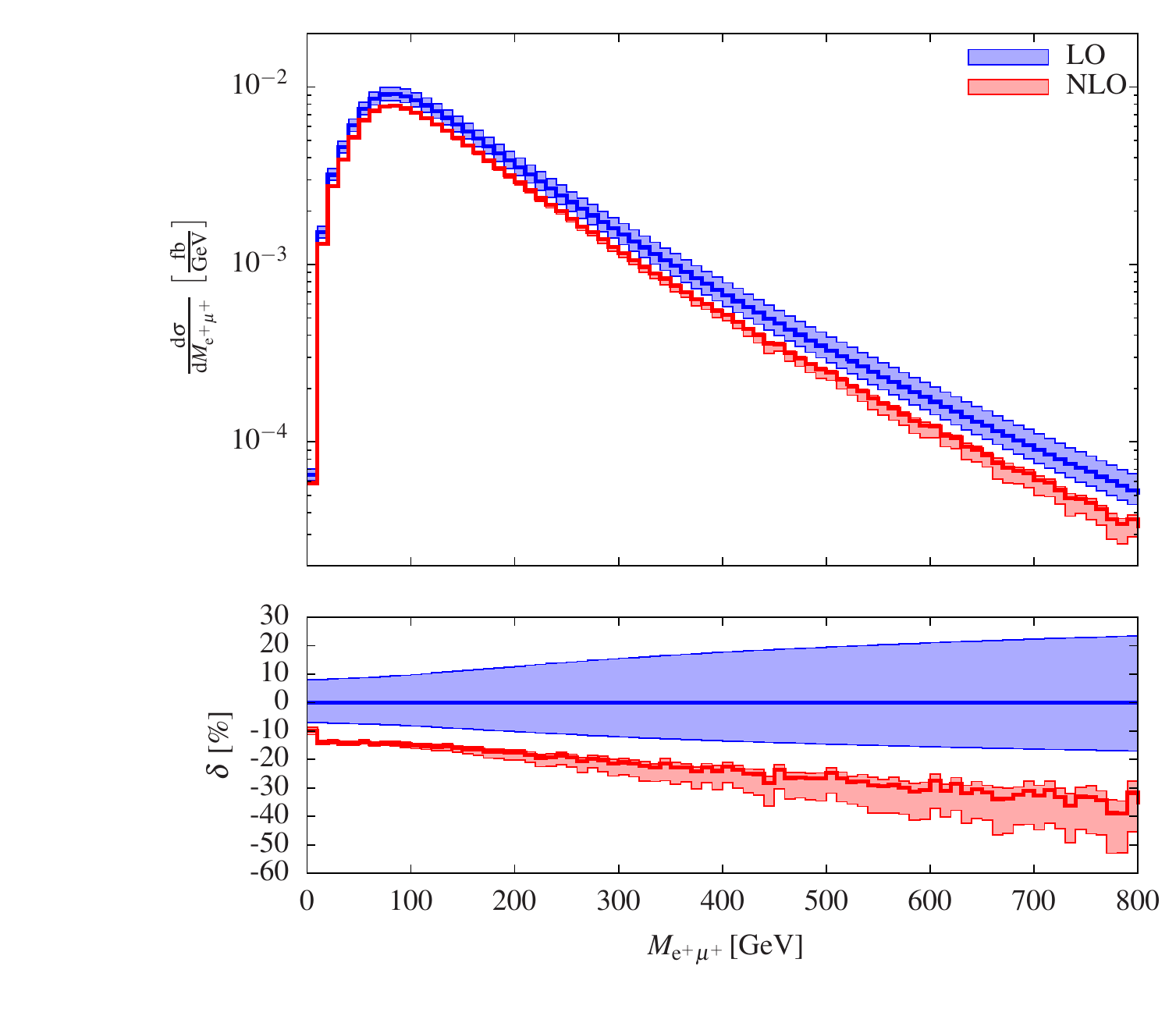}}
\caption{
These figures are taken from Ref.~\cite{Biedermann:2017bss}.
Differential distributions for the production of same-sign W bosons at the LHC.
The left plot shows the various LO and NLO contributions for the transverse momentum of the hardest jet.
In the right plot, all LO and NLO contributions are combined together.
}
\label{vbs}
\end{center}
\end{figure}

In Fig.~\ref{vbs} the full NLO predictions for same-sign W production are shown separately (left) and combined all together at LO and NLO (right).
The most interesting aspect here is the large EW corrections ($-13\%$ of the total LO fiducial cross section).
This is actually an intrinsic feature of vector-boson scattering at the LHC \cite{Biedermann:2016yds} that has been confirmed for almost all other signatures \cite{Denner:2019tmn,Chiesa:2019ulk,Denner:2020zit}.
In Ref.~\cite{Sirunyan:2020gyx}, such predictions have been compared to the CMS measurements for the same-sign W and the WZ channels.

They are reproduced in Table~\ref{cms}.
There, the LO predictions are obtained from the combination {\sc MadGraph5\_aMC@NLO}+{\sc PYTHIA} \cite{Alwall:2014hca,Sjostrand:2014zea}.
The NLO numbers are corrected with the NLO relative corrections of Refs.~\cite{Biedermann:2017bss,Denner:2019tmn} but applied only to the electroweak component.
Note that the uncertainty for the NLO numbers are obtained using 7-scales variation at LO.
The agreement between data and theory is in general good.

{\tiny 
\begin{table}
\centering
\begin{tabular}{cccc}
\hline
{Process} & {$\sigma \, \mathcal{B}$ (fb)} & Th. pred. LO (fb) & Th. pred. NLO (fb)   \\
 \hline
 \vspace{-0.3cm} \\
  {EW ${\rm W}{\rm W}$}     &  $3.98 \pm 0.45$            &  {$3.93 \pm 0.57$} &  {$3.31 \pm 0.47$} \\
{ {EW+QCD ${\rm W}{\rm W}$}} &  $4.42\pm 0.47$             &  {$4.34 \pm 0.69$} &  {$3.72 \pm 0.59$} \\
{EW ${\rm W}{\rm Z}$}    &  $1.81\pm 0.41$             &  {$1.41 \pm 0.21$} &  {$1.24 \pm 0.18$} \\
{{EW+QCD ${\rm W}{\rm Z}$}} &  $4.97\pm 0.46$             &  {$4.54 \pm 0.90$} &  {$4.36 \pm 0.88$} \\
{QCD ${\rm W}{\rm Z}$}    &  $3.15\pm 0.49$             &  {$3.12 \pm 0.70$} &  {$3.12 \pm 0.70$} \\
\hline
\end{tabular}
\caption{Comparison of fiducial cross sections between experimental measurements and theory predictions taken from Ref.~\cite{Sirunyan:2020gyx} for same-sign W and WZ production in association with two jets.}
\label{cms}
\end{table}
}

\acknowledgments

My research has received funding from the European Research Council (ERC) under the European Union's Horizon 2020 Research and Innovation Programme (grant agreement No. 683211).

\bibliographystyle{JHEP}
\bibliography{skeleton}

\end{document}